# Generation of broadband THz pulses in organic crystal OH1 at room temperature and 10 K

Andrei G. Stepanov[1], Clemens Ruchert[1], Julien Levallois[2], Christian Erny[1,3], and Christoph P. Hauri[1,3,*]

[1]*Paul Scherrer Institute, 5232 Villigen, Switzerland*
[2]*DPMC, Université de Genève, 24 Quai Ernest-Ansermet, CH-1211 Genève 4, Switzerland*
[3]*Ecole Polytechnique Federale de Lausanne, 1015 Lausanne, Switzerland*
*christoph.hauri@psi.ch*

**Abstract:** We studied the effects of cryogenic cooling of a 2-[3-(4-hydroxystyryl)-5, 5-dimethylcyclohex-2-enylidene] malononitrile (OH1) crystal on the generation of broadband THz pulses via collinear optical rectification of 1350 nm femtosecond laser pulses. Cooling of the OH1 crystal from room temperature to 10 K leads to a ~10% increase of the pump-to-THz energy conversion efficiency and a shift of the THz pulse spectra to a higher frequency range. Both effects are due the temperature variation of THz absorption and the refractive index of the OH1 crystal. This conclusion has been verified by temperature dependent measurements of the linear absorption in the THz frequency region.

## 1. Introduction

Generation of high energy ultrashort THz pulses [1] remains an active area of research due to the current and potential applications in nonlinear spectroscopy [2, 3], characterization of X-ray and electron ultrashort pulses [1, 4−6], and biomedical imaging [7]. Today, high energy near single-cycle THz pulses can be obtained at large scale accelerator facilities (up to 600 μJ [1]) or by optical rectification of femtosecond laser pulses with tilted pulse fronts (TPF) in lithium niobate (up to 125 μJ [8]) [9, 10]. The latter method represents a tabletop technique. In addition, the generation of ≥460 μJ single cycle pulses with an average frequency of 0.4 THz from solid density plasmas created by a tabletop multi-terawatt femtosecond laser system was recently reported [11]. The average frequency of TPF sources is typically below 1.5 THz. The strong increase in absorption in $LiNbO_3$ above 1.5 THz prohibits efficient THz generation at these frequencies. Cryogenic cooling of $LiNbO_3$ leads to a significant rise (by a factor of about 3.5) of the THz generation efficiency [10, 12, 13]. However, cooling does not allow substantially increasing the average frequency of generated THz pulses. To cover the frequency range between 1 and 10 THz, optical rectification in organic salt crystals (DAST, DSTNMS, OH1 etc.) have been proven to allow for a high pump-to-THz conversion efficiency [14−17]. Recently, the generation of 45 μJ near single cycle pulses with an average frequency of 2.1 THz and 15 μJ pulses with an average frequency of 2.65 THz has been demonstrated by this technique [14, 17]. It was found that in order to obtain a high pump-to-THz energy conversion efficiency (up to 2% [14]) in many organic salt crystals, the application of femtosecond laser pulses at 1.2–1.5 μm is required. The use of a $DAST/SiO_2$ multilayer structure for efficient generation of near single cycle pulses with an average frequency of 6 THz via collinear optical rectification of 800 nm femtosecond laser pulses was also proposed [18].

Previously it was found that cryogenic cooling of the widely used organic salt crystal DAST does not result in strong changes of its absorption and refractive indices [19] in the THz frequency range. Consequently, cooling does not increase THz generation efficiency or modify the THz pulse spectra noticeably [20]. However, the effect of cryogenic cooling on other organic salt crystals like DSTMS and OH1 that are used for high power THz generation [14−17] has not been studied before. In this paper, we report the cryogenic cooling effect on THz generation in the 2-[3-(4-hydroxystyryl)-5, 5-dimethylcyclohex-2-enylidene] malononitrile (OH1) crystal via collinear optical rectification of 1350 nm femtosecond laser pulses.

## 2. Experimental set-up

The experimental set-up used for studying the effect of cryogenic cooling on THz generation is show in Fig. 1. A 1 mm thin OH1 crystal with a 3 mm clear aperture was mounted on the cold finger of a closed-cycle helium cryostat equipped with polyethylene cyclic olefin copolymer (Topas) windows [21] for THz beam output. The crystal was irradiated by 1350 nm, 60 fs pulses coming from a high power OPA pumped by a TW power Ti:sapphire laser system [22]. The laser beam polarization was parallel to the crystallographic c-axis, which is the direction with the highest nonlinear susceptibility. In order to obtain a top-hat homogenous pump intensity distribution on the crystal surface, a broad NIR beam coming from an OPA was passed through a 3 mm aperture placed before the cryostat. The spectra of generated THz pulses were measured by a commercial Michelson interferometer with a Golay cell as a THz power detector.

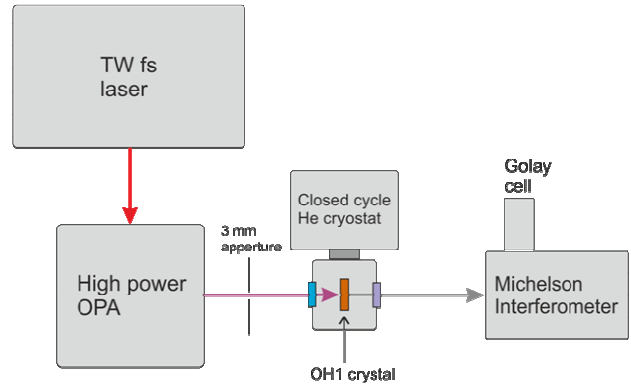

Fig.1. Experimental setup.

Linear absorption spectra of an OH1 crystal over a broad THz frequency range were measured at temperatures between 5 K and 300 K using a commercial FFT spectrometer.

## 3. Experiment results and discussion

Fig. 2 depicts the THz transmission spectra of an OH1 crystal with the optical axis oriented parallel to the THz wave polarization. The open cycle lines in Fig. 2 present experimental data obtained at different temperatures between 5 and 300 K, whereas the solid line are experiment data fits to a Drude-Lorentz model evaluated with a RefFIT program [23]. The fit parameters are shown in Tab. 1.

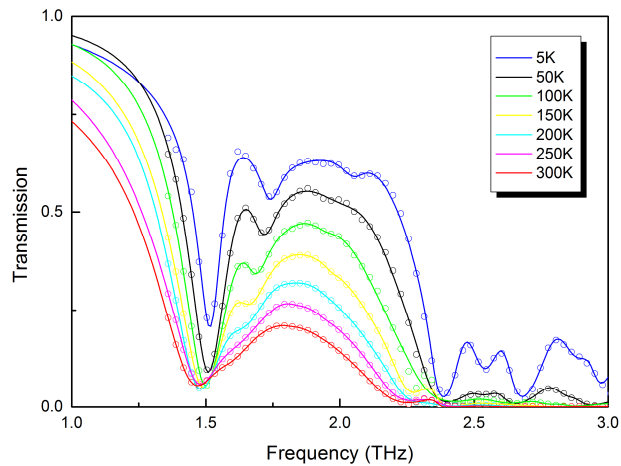

Fig. 2. THz transmission spectra of an OH1 crystal with optical axis oriented parallel to the THz wave polarization.

**Table 1. Fit parameters of the OH1 crystal transmission spectra (Fig. 2) to a Drude-Lorentz model model.**

| Temperature (K) | Oscillator frequency (THz) | Oscillator strength | Line Width (THz) | Temperature (K) | Oscillator frequency (THz) | Oscillator strength | Line width (THz) |
|---|---|---|---|---|---|---|---|
| 300 | 1.45 | 45.7 | 0.13 | 150 | 1.48 | 39.7 | 0.08 |
|  | 1.58 | 17.3 | 0.14 |  | 1.66 | 16.2 | 0.12 |
|  | 2.23 | 78.8 | 0.21 |  | 1.83 | 32.8 | 0.74 |
|  | 2.41 | 92.0 | 0.04 |  | 1.97 | 3.8 | 0.07 |
|  | 2.71 | 150.5 | 1.80 |  | 2.25 | 45.7 | 0.13 |
|  |  |  |  |  | 2.40 | 53.7 | 0.05 |
|  |  |  |  |  | 2.58 | 96.8 | 0.47 |
|  |  |  |  |  | 2.79 | 87.85 | 0.28 |
| 250 | 1.47 | 41.7 | 0.11 | 100 | 1.49 | 35.9 | 0.07 |
|  | 1.59 | 32.0 | 0.27 |  | 1.68 | 11.6 | 0.10 |
|  | 1.84 | 5.0 | 0.10 |  | 1.74 | 27.7 | 0.51 |
|  | 1.94 | 7.9 | 0.15 |  | 1.95 | 7.5 | 0.15 |
|  | 2.24 | 63.8 | 0.19 |  | 2.38 | 81.3 | 0.24 |
|  | 2.41 | 56.0 | 0.09 |  | 2.61 | 54.5 | 0.19 |
|  | 2.60 | 109.7 | 0.35 |  | 2.77 | 38.0 | 0.12 |
|  | 2.85 | 119.0 | 1.6 |  | 2.86 | 36.8 | 0.09 |
| 200 | 1.47 | 42.3 | 0.09 | 50 | 1.5 | 22.1 | 0.06 |
|  | 1.63 | 21.0 | 0.18 |  | 1.74 | 9.8 | 0.09 |
|  | 1.82 | 8.9 | 0.18 |  | 1.83 | 30.8 | 0.66 |
|  | 1.97 | 10.1 | 0.23 |  | 2.05 | 7.6 | 0.12 |
|  | 2.27 | 62.5 | 0.16 |  | 2.38 | 42.2 | 0.07 |
|  | 2.39 | 58.5 | 0.12 |  | 2.53 | 32.2 | 0.12 |
|  | 2.66 | 105.4 | 0.36 |  | 2.67 | 36.7 | 0.08 |
|  | 2.85 | 107.4 | 1.27 |  | 2.87 | 27.0 | 0.15 |
|  |  |  |  |  | 2.96 | 22.3 | 0.08 |

Fig. 3 (a, c) shows typical interferograms of the THz pulses generated at room temperature and 10 K via optical rectification of laser pump pulses with a moderate pulse energy (0.1 and 0.155 mJ). Corresponding FFT power spectra are shown in Fig. 3 (b, d).

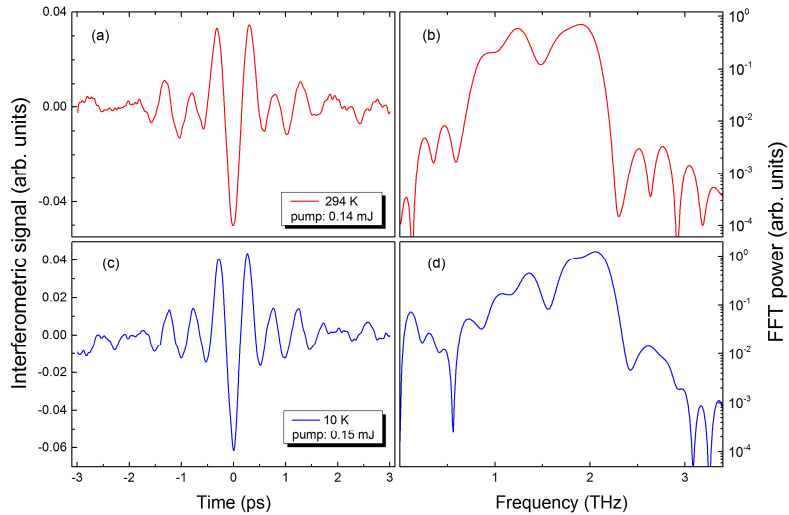

Fig. 3. Interferograms (a, c) and corresponding FFT power spectra (b, d) of THz pulses generated in an OH1 crystal at room temperature and 10 K via optical rectification of 1350 nm femtosecond with moderated pulse energies (0.14 and 0.155 mJ).

The spectra of THz pulses generated at different pump intensities at room temperature and 10 K are presented in Fig. 4 (b-d). The overall THz pulse spectra are in good agreement with the THz linear absorption spectra of the OH1 crystal shown in Fig. 4(a) and with previously reported spectra of THz pulses generated in this crystal [24]. The dip at about 1.5 THz in the pulse spectra corresponds to an absorption line at this frequency. Cryogenic cooling of the OH1 crystal leads to a power decrease and narrowing of this absorption line. Moreover, cooling results in an increase in the central line frequency. Such a thermal frequency shift was observed in a DAST crystal for the absorption line at 1.2 THz [18]. However, the absorption line at the central frequency of 3.4 THz in the DAST crystal shows neither a thermal frequency shift nor a narrowing with temperature [20]. Despite a decrease in THz absorption in the OH1 crystal with temperature around 1.5–1.7 THz, we did not observed an increase in THz pulse spectral power in this frequency range. This phenomenon probably originates from the phase mismatch appearing at these frequencies upon cooling, as the group phase refractive index at these frequencies decreases relative to the phase refractive index at 1350 nm. The same effect leads to a decrease in THz pulse power in the frequency range of 0.5–1.3 THz. The decrease in THz pulse spectral power above 2 THz at room temperature results from a rise in absorption and phase mismatch related to a strong and broad absorption band centered at about 3 THz. An increase in THz pulse power in the 2−2.2 THz spectral range observed upon cooling is determined by two effects: a decrease in THz absorption and better phase-matching at these frequencies due to a narrowing of the absorption band at 3 THz. In previous experiments at room temperature [24], a small increase in THz spectral power in this frequency range could be achieved by improved phase matching by tuning the pump laser wavelength from 1400 nm to 1250 nm. We assume that the same effect leads to an increase of the small peak at 0.1 THz for the cooled crystal (Fig. 4 (b-c)).

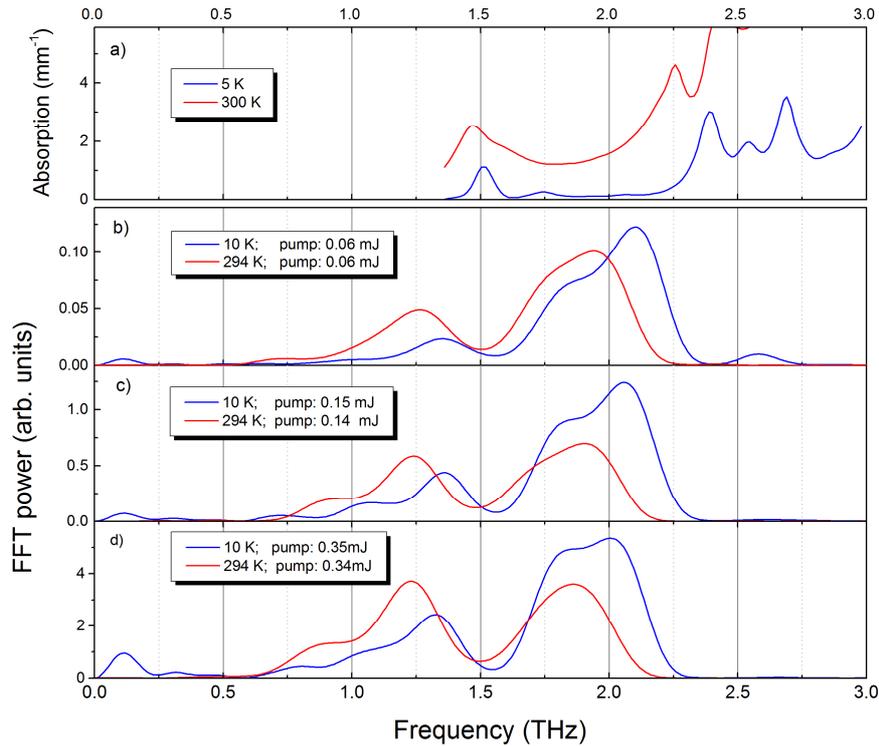

Fig.4. Linear absorption THz spectra of the OH1 crystal used in this study (a). Spectra of THz pulses generated in this crystal by optical rectification of 1350 nm laser pulses with different pulse energies at room temperature and 10 K (b-d).

In addition, the low frequency part of the THz pulse spectra (centered at 1.25 THz at room temperature) rises faster than the high frequency part (centered at 1.8 THz at room temperature) with pump intensity. In particular, this effect is noticeable at room temperature (Fig. 4 (b-d)). This effect can appear due to a saturation that starts earlier at high frequencies than at low frequencies because of the increase in THz generation efficiency with the square of the frequency (second order nonlinear processes). Moreover, this phenomenon can be thought of as a manifestation of the cascaded character of the lower frequency peak generation.

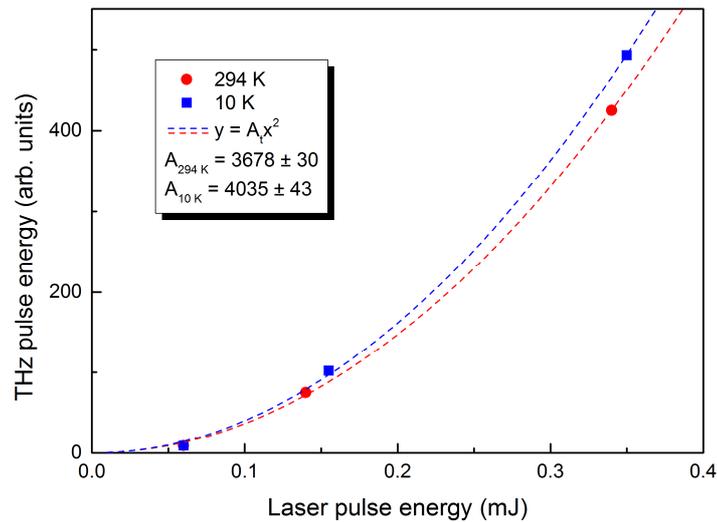

Fig. 5. THz pulse energy as a function of the laser pulse energy measured at room temperature and 10 K.

Fig. 5 represents the THz pulse energy as a function of laser pulse energy measured at room temperature (circles) and 10 K (squares). Arbitrary values of THz pulse energy were obtained by integrating the THz pulse spectra shown in Fig. 4 (b-d). The measured points correspond well to the pump pulse energy squared (dashed lines in Fig. 4). This square law dependence implies that saturation is negligible in the investigated intensity range. Cooling of the OH1 crystal from room temperature to 10 K increases the THz pulse energy by 10%. This increase is a relatively small improvement as compared to the THz generation efficiency rise observed in lithium niobate at cryogenic temperatures. The difference is probably related to the weaker phonon-phonon coupling in organic crystals, which leads to a relatively weak THz absorption and thus has a relatively limited potential to change in absolute value with temperature.

### 4. Conclusion

We have experimentally shown that the generation efficiency of ultrashort THz pulses via collinear optical rectification of 1350 nm femtosecond laser pulses in OH1 can be increased by 10% through cryogenic cooling. Cooling also leads to a shift in the THz pulse spectrum towards the higher frequency range.

**Acknowledgement**

This work was partially supported by the Swiss National Science Foundation grant PP00P2_128493 and by NFS (grant no 51NF40-144615) in the framework of National Center of Competence in Research (NCCR-MUST) and SwissFEL .